# SECURITY AND NON-REPUDIATION FOR VOICE-OVER-IP CONVERSATIONS

**Christian Hett, Nicolai Kuntze, and Andreas U. Schmidt**

Fraunhofer-Insitute for Secure Information Technology SIT

Rheinstrasse 75

64295 Darmstadt, Germany

Phone: +49 6151 869 60227

Fax: +49 6151 869 224

{christian.hett,nicolai.kuntze,andreas.u.schmidt}@sit.fraunhofer.de

ABSTRACT

We present a concept to achieve non-repudiation for natural language conversations by electronically signing packet-based, digital, voice communication. Signing a VoIP-based conversation means to protect the integrity and authenticity of the bidirectional data stream and its temporal sequence which together establish the security context of the communication. Our concept is conceptually close to the protocols that embody VoIP and provides a high level of inherent security. It enables signatures over voice as true declarations of will, in principle between unacquainted speakers. We point to trusted computing enabled devices as possible trusted signature terminals for voice communication.

# SECURITY AND NON-REPUDIATION FOR VOICE-OVER-IP CONVERSATIONS

## 1  INTRODUCTION

The latest successful example for the ever ongoing convergence of information technologies is internet based telephony, transporting voice over the internet protocol (VoIP). Analysts estimate a rate of growth in a range of 20% to 45% per annual, expecting that VoIP will carry more the fifty percent of business voice traffic (UK) in a few years [1]. The success of VoIP will not be limited to cable networks, convergent speech and data transmission will affect next generation mobile networks as well. The new technology raises, however, some security issues. For eavesdropping traditional, switched analogue or digital phone calls, an attacker has to get physical access to the transport medium. Digital networks are generally more amenable to attacks. This holds already for ISDN and to a yet greater extent for IP networks. Efforts to add security features to VoIP products are generally insufficient, though proposals exist for privacy protection. Secure VoIP protocols, using cryptographic protection of a call, would even be at an advantage compared to traditional telephony systems. Protocols like SRTP can provide end-to-end security to phone calls, making them independent from the security of the transport medium and the communication provider [2]. This can also be beneficial in view of the high security requirements for wireless phones.

On the other hand voice conversations provide inherent evidentiary value due to the possibility of forensic evaluation and analysis of the contained biometric data, e.g., as an independent means of speaker identification [3, 4]. Methods for the latter are advanced [5], obtaining to recorded voice communication a rather high probative force, e.g., in a court of law. In comparison to other digital media, e.g., text documents, specific features of voice communication can be viewed as contributing to security. The medium of communication here consists in a linearly time-based full duplex channel enabling inter- and transactivity [6]. In particular, interactivity enables the partners to make further enquiries in case of insufficient understanding. Furthermore, digital voice communication offers a rather high reliability and quality of service, leading generally to a higher understandability of VoIP communication in comparison with its analogue predecessors [7, 8]. The mentioned properties mitigate to some extent problems to which digital documents are usually prone, e.g., misinterpretations due to misrepresentation, lack of uniqueness of presentation, and inadvertent or malicious hiding of content.

Therefore, we depart from the basic security aspects of VoIP communication and view conversations on a transactional level between caller and callee. The top-level category of protection targets that we consider is *non-repudiation of conversations*. Three tasks of ascending complexity are addressed in the present work:

1) Protection of the integrity of voice conversations. Protecting a (recorded, digital) voice conversation from falsification and tampering with is different from protecting the integrity of other digital data due to the relevance of the temporal context. In particular, packet ordering and loss have to be considered properly, and a creation time must be assigned to each conversation.

2) Authentication of speakers. An initial authentication of caller and callee together with the inherent biometric authenticity of voice is the basic approach to this problem. While it could be resolved in principle solely on the transport layer, it is advantageous to combine it with the methods of 1) to obtain proof that a (recorded) conversation was carried out completely from the authenticated devices. It has to be noted that each authentication of a speaker requires trust in the devices used by the communication parties.

3) Electronic signatures over voice conversations. Building on 1) and 2) it is possible to achieve, for voice conversations, the level of non-repudiation provided by electronic signatures over digital documents, i.e., an expression of will. For this, the aforementioned tasks must be complemented by a proof of possession of a trustworthy signature token and device, and the intention to sign.

We present theoretical and technological concepts for each of the tasks 1) – 3) and describe their realisation in a demonstration environment. The existing VoIP infrastructures are largely unaffected by our concepts through a seamless and efficient integration in the SIP [9] and RTP protocols. On the other hand, the three tasks pose increasing technical requirements on the part of the involved devices.

The paper is organised as follows. Section 2 lists requirements for a technical solution to the three tasks at hand both form a security and an efficiency viewpoint. Section 3 introduces the application scenario in which digital signatures over voice communication are realised and explains the base concept for the resolution of our tasks. A detailed technical implementation is described in Section 4, and Section 5 contains a security analysis of it.

Task 1) is resolved in a stand-alone way, without any change in devices or transport methods. The solution concept rests on cryptographic secrets created at the initiation of a call and their perpetuation throughout the call by a cryptographic chaining method. As a key issue, stability, quality of service, and necessary fragility to prevent attacks must be balanced with each other. As a main application of the concept we view a secured archive for VoIP conversations which yields tamper-resilience and in consequence evidentiary value by far exceeding that of traditional tape archives. On the other hand, tasks 2) and 3) need additions on the used devices ranging from the inclusion of authentication software and the roll-out of pertinent data, to the fulfilment of high security requirements for signature terminals. Though difficult in general, the latter task is simplified since only one channel (audio) carries the signed information. We propose a concept employing methods of trusted computing to turn a VoIP capable handset into a trustworthy signature platform in Section 6.

The combined benefits of the technology developed here amount to a new paradigm for non-repudiation of digital data. The combination of integrity of recorded conversations, security about the identity of dialogue partners, and finally expressions of will embodied in signatures enables legally binding verbal contracts between unacquainted persons. Perfect security could only be achieved by recording the conversation and afterwards replay the recording to the participants. This solution raises various problems as the recording is unprotected during the recording and the replay has to be protected as well. The practicality of such an approach is also doubtful.

In the context of recording a conversation between two parties it is important to stress that all involved parties have to agree to the recording which would otherwise the be void. This explicit or implicit agreement is to be recorded as well. State of the art is a short message addressing the partner or it is part of the contract between the communication parties (e.g. in the home banking area). In more advanced scenarios [10] request and agreement can be signalled using the abilities of user interaction and user defined policies. One example is [11] where privacy levels can be defined and therefore should be evaluated by the respective implementation.

## 2    REQUIREMENTS FOR VOIP NON-REPUDIATION

The central requirements for achieving non-repudiation by signing VoIP are, of course, related to security. Of the well known trinity – Confidentiality, Integrity, and Availability – of information security requirements, **integrity** is the central one to achieve non-repudiation for digital, packet-based, natural language communication. It needs to be assured that a communication was not changed at any point in time, be it during transmission or later. Furthermore, integrity comprises as well the integrity of any relevant meta-data created or used during a call, in particular the data that authenticates the communication partners, or at least the partner exerting the signature over the communication.

Due to the special features of voice communication, i.e., a bidirectional, full-duplex interactive conversation, only both channels together provide the necessary context to fully understand the content of the conversation and to make use of the inherent security that interweaved natural language conversations provide. To ensure that parts of the talk are not exchanged with other parts, replaced by injections, or cut out, the envisaged system needs to ensure what we call **cohesion**. That is, the temporal sequencing of the communication and its direction is data the integrity of which needs to be protected in a way that makes later tampering practically unfeasible, i.e., by sufficiently strong cryptographic methods. Cohesion as a feature related to time entails a subsidiary requirement, namely the secure **assignment of a temporal context** to a conversation. Each conversation has to be reliably associated with a certain time, which must be as close as possible to the conversation's start and the initiation of the signing (note that assignment of a signing time is a legal requirement for qualified electronic signatures according to the European Signature Directive and pertinent national regulations). Drift of the time base should be mitigated during a signed conversation. Finally, cohesion also refers to qualitative aspects of the communication channel. A signatory is well advised not to sign a document which is illegible or ambiguous. In the digital domain this relates to the presentation problem for electronically signed data [12]. In analogy, the **quality of the VoIP channel** must be maintained to a level that ensures understandability to both partners during the time span in which the conversation is signed.

To ensure later availability of a signed conversation is a trivial requirement for the receiver to be able to make use of it as evidence. The secure archiving of VoIP communication is not in the scope of the present paper. A solution using the methods that we exhibit here is contained in [13]. Considerations of long-term archiving aspects for signed digital data can be found in [14, 15]. Confidentiality is also out of the scope of the present contribution, since it can be resolved at the transport layer.

Further requirements regard the efficiency of the system design and implementation. First, it is highly desirable, both from a security as well as an efficiency viewpoint, to sign and secure the VoIP conversation as "close" as possible to its transmission, and conceptually close to the actual VoIP stream. Simplicity of the implementation should minimise the effect on existing systems and infrastructures, e.g., client-side requirements should be minimised. A tight integration is required to enable the utilisation of existing infrastructures without or with only minor changes. An efficient use of memory, bandwidth, storage space, and computational resources can be achieved by basic conceptual design decisions. Furthermore, scalability of the concept to a large number of concurrent calls is a necessity in real business environments. This means that centralised signature creation infrastructures should be avoided if possible. Finally, any architecture that copes with VoIP security issues needs to appropriately take packet loss into account, in particular in view of the cohesion requirement. Packet loss leads to modifications of the conversation perceived by the receiver. Loosing a single packet leads to a loss of only some ms. On the other hand the quality of VoIP connections is very high so that loosing a single packet is not very likely. In our tests we haven't seen packet loss at all (However, complete drops of the communication are not so infrequent, yet they are covered well by our approach).

## 3  SIGNING SCENARIO AND BASE CONCEPT

The main scenario of this paper is shown in Figure 1 and is a bidirectional interactive conversation between two parties A and B. A wants to sign the conversation and release it to B as a declaration of his will, i.e., a commitment in the sense of a signed offer. In effect, A wants or is required to make sure that the conversation between A und B provides non-repudiation, in any case he expresses the explicit will to make non-reputable statements or stipulations in the call. For that, A possesses a digital certificate for a signature public/private key pair. B and any third party to whom the signed conversation is presented as evidence, are assumed to be able to verify the certificate of A and any data signed with the associated private key, i.e., we assume a PKI structure in the background. A signs the complete call including both channels comprising everything that B says. B stores the signed conversation in a secure archive of his choosing. B can later proof to third parties or court that the call happened and had the claimed contents. If B fails to store the conversation in an archive or deletes it, A can deny that the call ever happened. Cohesion is ensured by the used protocol and procedure, so that the full bidirectional interactive dialogue is continuously signed.

As part of the implementation, the ability to record and archive a conversation on the part of B had to be developed. While the main scenario requires changes to the terminal equipment on both sides for signing and interchange of signed data, we were also able to apply similar methods to a pure archiving scenario [13]: A component VSec, implanting the integrity of the archived conversation, could be placed anywhere in the way from A to B as long as any part of the communications was carried out over SIP/RTP. Thus the archive functionality can be added to the corporate VoIP telephone system, into the devices used by A and B or integrated in any other point which has access to all exchanged packages of the communication. Using this, party B can source out the archive from VSec. The communication between the component VSec which listens and intercepts the conversations and the archive itself is secured. These methods are applicable in the present scenario, and are particularly interesting considering the limited storage space of mobile phones.

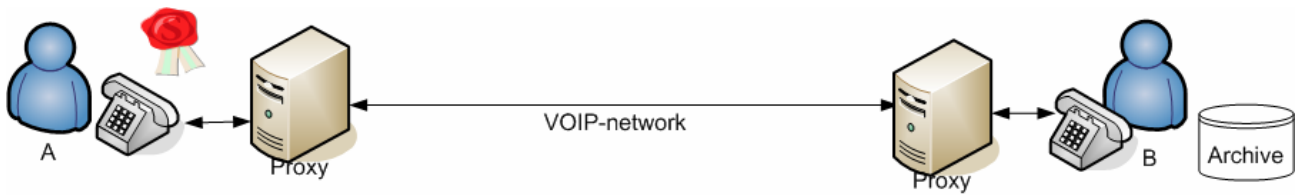

*Figure 1: The main scenario where A possesses a digital certificate and provides non-repudiation for the whole conversations and B archives it for later possession.*

An extension to this concept is that both parties could sign the conversation. This is sensible in the form of parallel signatures, i.e., each party signs their own utterances and what the other party said and both parties archive it. This slight variation provides mutual non-repudiation.

The signing scenario is implemented in a flexible demonstrator placed as a proxy between the internet and the soft-phone or VoIP hardware phone of party A and between the internet and the phone of party B. The demonstrator manages certificates and signing of the voice data and also archives the conversation if used by party B. The proxies shown in Figure 1 are an implementation detail as our demonstrator does not integrate with a specific VoIP client directly, but is located on the same computer or in the same network and can be used with any SIP phone.

## 4    IMPLEMENTATION

The implementation of the singing scenario is based on the SIP/RTP protocol, but the method could also be applied to other protocols like the Inter-Asterisk-protocol IAX [16] or the well established H.323. The signing protocol extends SIP/RTP [17] in a compatible way to transport signatures and acknowledgments of signatures. Instead of modifying a particular type of (soft-)phone, we chose to implement this as a proxy that intercepts the SIP call signalling and also the RTP audio (and video) streams. A strength of the technique is that it does not modify or in any way delay the transported audio stream. Instead signatures are transported sparely and separately from the audio stream.

### 4.1    Achieving Cohesion by Interval Chaining

As stated in section 2 it is important that the implementation is efficient and scalable. Therefore solutions like signing every single packet with a signature algorithm like RSA is neither efficient with respect to bandwidth and storage capacities, nor sufficient to protect the full conversations.

Signing each packet alone easily uses more than 100 bytes to store a signature of an RTP packet with only 44 bytes of sampled audio and is computationally expensive. Therefore we introduce the central concept of **intervals** and **interval signatures**.

Each party collects packets in intervals of adjustable length, e.g., one second. Time based intervals pose certain problems as it is hard to predict size of the required caches or suspension of the timer. So, other solutions to determine the duration of a interval can rest on the sent and received amount of packages. For the sake of simplicity we focus in this paper on timer based events. Every second the collected packets are sorted by sequence number and their hashes are assembled in a data-structure with additional meta-information like direction, sequence numbers and time. This small data-structure is then signed with a conventional signing algorithm like RSA, using the private key of party A. They are then sent to B who then stores them together with the collected RTP packets he actually received. Note that the full packets are transported only once as in a normal RTP stream. Therefore bandwidth as well as CPU time is drastically saved making the whole method applicable in the first place (computing hashes is much less expensive compared to RSA signatures, especially on mobile phones with limited processor-speed).

As a side note, it would be possible to further reduce the bandwidth usage since the sequence number of the packet is enough for B to reconstruct the hash, rendering the transmission of hashes unnecessary. Transmitting only packet numbers would bear the cost of additional consistency checks on the part of B. We do not discuss this detail further because the result would be the same but presentation would be more complex than signatures built only on actual data. It is also important to stress that the signature for a complete interval is broken if any of the hash values becomes invalid, namely, if any bit of the signature or in the associated and stored RTP packets is changed, or if any packet is missing. This is in strong contrast to the technique of stream signatures presented in [18] where the authors show methods of signing unidirectional broadcast traffic in a way that makes it possible to still check a signature if packets are lost in transmission. Because lost packets are very common in VoIP scenarios, but on the other hand lost packets are also potential attack vectors to our approach, we use a different technique to deal with packet loss as detailed in Section 4.3.

Signed intervals alone do not ensure cohesion. An attacker could exchange parts of the conversation or cut them out. Therefore we make use of hash chains: Every interval contains, embedded in its metadata, a hash of the last interval including its signature. In this way signatures and hashes are interleaved ensuring that there is a continuous stream of signatures building an unbreakable chain. The chaining of intervals is further extended to factor into the bidirectional nature of the call. Both channels are interwoven and the chaining applies to both channels. An interval of packets from the channel A→B contains a hash of the last signed interval from the channel B→A and so on. This way cohesion is strongly secured by RSA signatures.

### 4.2 Resulting Signed Data Format

As non-repudiation of calls is only meaningful if the party who is interested in using a conversation as evidence – in this case party B – the signed conversation must be recorded. Special emphasis must be put on the format in which the calls are stored, i.e. the final outcome of the signing protocol. All intervals of a call are simply stored continuously by the proxy software of party B. In general additional timestamps (as can be seen in the start chunk in Figure 2) may help pinpointing the exact start and duration of the call.

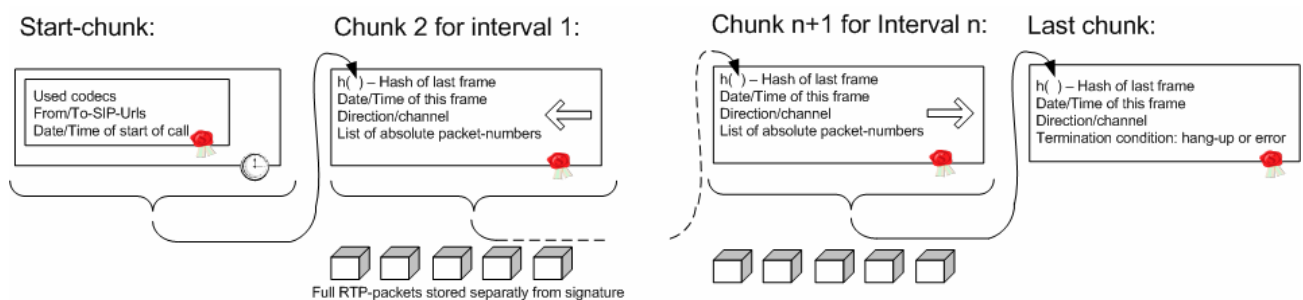

*Figure 2: The hash-chain of signed interval-packages in our data format. Note that B stores the collected packets next to A's signature for each frame.*

The format is shown in Figure 2 and is a simple chunk-based continuous data format starting with an initial chunk containing meta-data (SIP URIs of caller and callee, date and time of start of call and mapping of codecs to RTP payload fields). After that signed intervals are stored as they are produced, reducing working memory requirements to the order of magnitude of one interval. At the end of the call (either by normal hang-up or by policy termination as described in Section 4.4) a special chunk is added which contains time and the reason for the termination of the call.

For each interval the associated chunk contains the collected RTP packages of this interval, and the following signed data: The direction/channel of the interval (from A to B or vice versa), the date and time, the list of absolute package sequence numbers and the hashes of each considered package. This data is embedded in a PKCS#7 signed envelope container. The signing is done by party A using his private key. Only the first PKCS#7 envelope needs to store the whole certificate chain, all other envelopes do not need to store any certificates.

### 4.3 Signing Protocol

In this section we describe the protocol that is used to transport acknowledgments, signalling, and signatures for the normal RTP packets containing multi-media frames. The protocol is a simple compatible extension to the existing SIP/SDP/RTP system and can be transported easily by the existing infrastructure. This includes the typical solutions for common problems arising form NAT routers, like firewall-hole-punching and TURN servers [19].

In the signalling phase of a SIP call the description of the call parameters, including IP-addresses, ports, and multi-media codecs and their parameters, are transported in the body of SIP-messages using the simple SDP protocol [20]. SDP is able to negotiate the transport characteristics of more than one media per call, e.g., an audio stream, a video stream, and in addition an application-stream for, e.g., whiteboard data. We use such an additional application stream for transportation of signature data. Note that this stream is unreliable and datagram-based and not a reliable stream like TCP because it is based on RTP which runs on UDP.

For the initial signalling of the fact that someone is calling who has signature capabilities and is willing to sign the conversation we use the k-value of the SDP-protocol which is used for securing the call using SRTP (many other variants may exist). Here, A transmits essentially the same data as in the first chunk in Figure 2, signed using his private key.

To further describe our protocol it must be stressed that even though both channels of the duplex conversation need to be signed to provide cohesion, in our base scenario still only party A needs to have a private key for signing. A has to sign both directions of the conversation. Accordingly the signing protocol differs for both channel directions which are described in the next two subsections.

#### 4.3.1 Signing the Channel from A to B

For the channel direction A→B, A could simply send (every second) an interval containing the hash-values of all his RTP-packets of this interval. If any packet-loss occurs – which is normal in current networks – B wouldn't be able to provide the missing packets as evidence. In this case his whole archive of the call would be deprived of probative force. Therefore in our protocol B transmits a list of all sequence numbers of the packets he received during an interval to A, who will then send the requested signature that covers exactly the received packets.

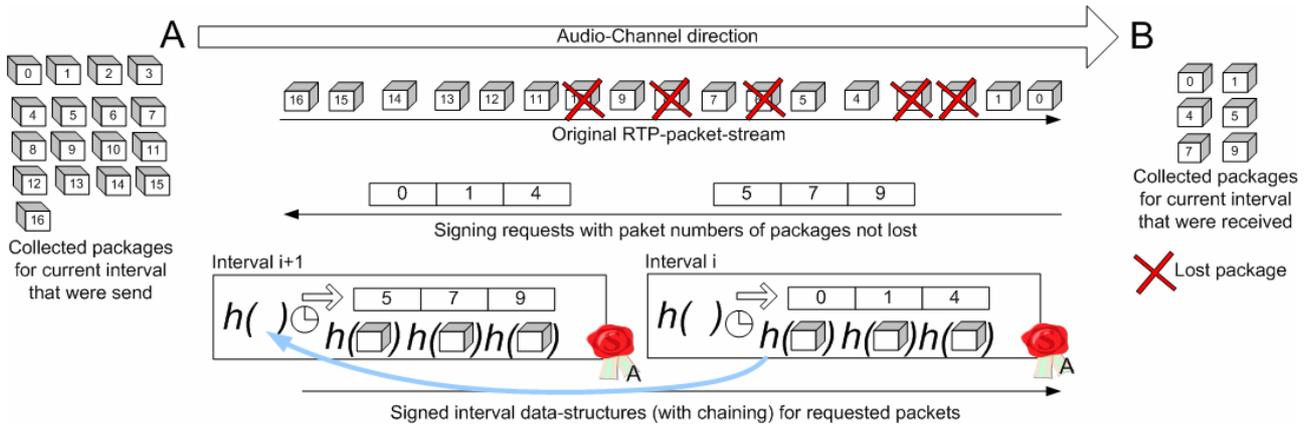

*Figure 3: Schematics of signing protocol for the direction A→B.*

More precisely both A and B continuously collect all packets for the channel A→B in a small buffer, as indicated in the left- and rightmost parts of Figure 3. Whenever B's interval timer expires he will send a list of packet sequence numbers that were collected during that interval to A and starts a new interval. The transfer of the package sequence numbers can be performed asynchronously, while the conversation is going on. A will then create the hash values out of the collected packets and create and sign an interval data structure as described in Section 4.2 and transmit it to B. B then checks the signature and hashes and stores this together with the collected packets and drops the collection of packets from memory. A can drop his complete packet collection from memory one iteration later, because if the communication that is described in this paragraph fails, then B will retransmit its packet list until A has successfully transmitted the interval signature. While waiting for the response, the interval timer of B is temporarily suspended as it can only fire after an interval signature was successfully transmitted. During this suspension of the interval timer the signing process does not stop. The timer is restarted if all data of the last interval is transmitted. If the transfer of this data needs too much time (e.g. longer than the interval length) the system should react as described in section 5 and consider this as a quality of service under-run. The whole process is sketched in Figure 3.

To mitigate several attacks, extensive checks are enforced on both parties to make sure that, e.g., B is not able to pretend that there was packet loss while the speakers never noticed this in their conversation. B could use this to selectively mute unfavourable sections of the conversation. The policies for these checks are described in Section 5.

### 4.3.2 Signing the Channel from B to A

The communication for signatures for the channel B→A is a little bit different. Again A and B both collect all packets for the current interval. But this time A decides on the precise point in time at which a new interval starts. Then A takes the collected and sorted packets which are exactly the packages that were not lost in the transmission from B to A, creates hashes over them, and sends them in a signed interval data structure to B. When B receives this it has to send a short acknowledgement package, containing the number of the interval, to A. If A does not receive this before a timeout occurs, he will resend the signature package. Until the acknowledgement package is not received, A will not start a new interval and the interval timer is temporarily suspended. We omit a graphical representation due to lack of space.

B also has to check the received signature package as described in Section 5, e.g., B has to check the quality of service of the conversation in direction B→A that is signed by A. This enables B to detect for instance whether A wants to conceal (on protocol/signature level) portions of B's utterances in the signed conversation. If A refuses to sign enough of what B sends, B should consider the conversation tampered with.

## 5 SECURITY CONSIDERATIONS

In a non-repudiation scenario, the most important attacks are those affecting the actually signed contents. Those attacks are mostly exerted by the signatory A, or the receiver of the signed content, B. Section 5.1 sketches some attacks by A and B on the integrity and cohesion of the signed conversation and Section 5.2 describes the countermeasures, in the form of security policies and checks, which can be effectively used to mitigate them.

We do not provide an exhaustive security analysis of the overall voice signing system. A more thorough treatment for the special case of a secure VoIP archive is contained in [13]. Furthermore, the described protocol does not contain any measures for confidentiality of the conversations. For this we refer to SRTP [21] which solves this problem very well. SRTP also ensures integrity on the packet level.

### 5.1 Some Possible Attacks

The protocol presented in Section 4.3 is able to deal with packet loss and will only sign packages that were not lost during the transportation. But this opens the door to a whole class of attacks. Party B is always able to delete the whole recorded conversation. But B could also try to change the contents of the conversation. He might be interested in omitting unfavourable sections of the conversations and bringing the remaining sections to court as evidence. Even worse he could rearrange sections to give them a different meaning. In spite of the countermeasures below which mitigate most of these attacks, the possibility for B to cut an arbitrary final section of a call remains.

Furthermore, either A or B could simulate packet loss. If other kinds of disturbances apart from packet loss occur in a normal conversation, it is natural to ask the partner to repeat that sentence. But assume that packets arrive at B, B decodes them, and hears them clearly, and therefore knows what was talked about and can continue the natural language dialogue. Now, if B lies about his packet loss and sends a list of very few sequence numbers back to A, then A would only provide a signature for these packets. A would not notice this and has no possibility to correct the mutilated part of his (incorrectly signed) utterance. If the recorded conversation is later played back, the packet loss would leave only few audio frames and A would have signed an unfavourable piece of conversation. The same mutilation attack could be used by A who could pretend that only parts of B's communication arrived.

An attacker or B could try to submit two streams of audio: An early one that is heard and understood by A and also a second one to invalidate the recorded signed conversation or to embed other words in it. A would then sign a part of an audio stream without ever hearing it.

### 5.2 Security Policy and Checks

As B has an interest in using the recorded signed conversation as evidence, he has to check whether the signatures from A are valid. In particular he can check that the certificate actually is issued to A, e.g., for A's SIP-URI, that it is trusted, that each hash-value is correct, and that the signatures are all valid and correctly hash-chained. Verification policies for A's signature depend on the level of speaker authentication that is desired. All these checks can be done in real time during the conversation so that any problem can be detected in a short time span. This effectively resolves the second of our three tasks proposed in the introduction.

Attacks by B on integrity and cohesion are suppressed by the hash chain and/or the hashes of individual packets. A changed bit is quickly detected.

To mitigate attacks aiming at mutilating a conversation as described above, quality of service (QoS) is permanently monitored. When the packet loss is above a configurable threshold (we use 1% in our demonstrator) the software notices this. It is now a matter of policy how to deal with this

QoS under-run. It could be ignored, the users could be notified while continuing the signature, the signing could be aborted, or the call could be terminated. The first two options open the path for described attacks. We favour termination of the call as this the option for maximum security and the QoS threshold is seldom reached without a breakdown of the connection anyway due to insufficient understandability or software timeouts.

This inherent feature of fragility of the signature scheme deserves some extra discussion. The condition to break if a QoS under-run occurs provides a Sollbruchstelle (predetermined break point) for the probative value of the signed communication. In contrast, most other schemes for securing the integrity of streamed data, e.g., the signing method of [18] aim at loss-tolerance, for instance allowing for the verification of the stream signature with some probability, even in the presence of intermittent packet loss. We argue that for the probative value of inter-personal, natural language communication, the former behaviour is advantageous. A signed call with an intermediate one-second gap can always give rise to speculations over alternatives of filling the gap, which are restricted by syntax and grammar, but can lead to different semantics. Using this, a clever and manipulative attacker could delete parts of the communication claiming with some credibility that the remnants have another meaning than intended by the communication partner(s). But if the contents of a conversation after such an intentional deletion are unverifiable and thus cannot be used to prove anything, this kind of attack is effectively impeded.

The central measures for providing cohesion and the integrity of the signature process are:

- A replay window (as recommended for SRTP) avoids the attack by inserting a second stream, since duplicate packets are detected and dropped. Depending on policy detection of duplicates may as well be a reason for termination of the call. Introduction of a replay window in turn makes availability attacks possible, which could be mitigated with HMAC integrity checks as in SRTP. The replay window also assists with sorting because the internet allows RTP packets to arrive out of order, but the intervals demand strictly monotonic sequence numbers: Packets are not passed to the package-collection component until they leave the window.

- The sequence numbers in the RTP packets are checked for the packet loss ratio and for a correct, monotonic increasing sequence. This happens on both sides and not only for incoming packets, but also for signing requests from B to A. A can refuse to continue signing the conversation if too few of his packets arrive at B so that he can still be understood.

- The timestamp that every RTP packet contains is checked against the wall clock (ideally, a trusted time source). If the discrepancy is too high, then this is a policy violation.

- The borders of the intervals are checked against the respectively next interval, ensuring that there is no gap larger than few packets.

- The intervals of both channels are checked to match in time so that they cannot drift away from each other and break cohesion.

With these policies we presented a consistent approach to fulfil the requirements of integrity and cohesion. In view of scalability and efficiency, by adjusting the interval duration a trade-off between required computational power and the unprotected time can be adjusted. We believe that our default time of twice the interval time of 1 second is already sufficient to make it very hard to change the meaning of a signed conversation. An attacker should not be able to forge more than two seconds at the very end of the conversation before the call is terminated. It should be noted that all described checks can as well be performed in a forensic evaluation, since all the data on which checks are performed is signed and secured.

## 6 TOWARDS A TRUSTED VOICE SIGNATURE TERMINAL

As mentioned above signed voice transactions are building up on the ability to 1) authenticate the communication party and 2) to protect the integrity of voice conversations. Considering existent authentication schemes achieving 1) requires a strong protection of the authentication secrets and, especially regarding biometry, a provable trust in the authentication hardware. To accomplish 2) we have introduced the VoIP signing protocol as a building block. To finally arrive at a trustworthy voice signature terminal which resolves also task 3) of the Introduction we need to testify that the hard- and software is meeting the established requirements to be considered as trustworthy input und signature creation units [22].

Shielding of secrets and testifying a state to a third party can be achieved by using the principles of Trusted Computing (TC) as defined by the Trusted Computing Group (TCG). The TCG defines the design principles of trusted platforms. As the root of trust TCG has introduced the Trusted Platform Module (TPM), which is a specialised hardware token which is integrated in a trusted platform. This TPM allows creating, storing, and using keys in a secure way by offering shielded capabilities which are protecting the private portion of a key. As a second relevant feature the TPM stores and shields trust measurements. These trust measurements can be used to perform a remote attestation where a third party receives these trust measurements signed by the TPM with the Attestation Identity Key (AIK). Based on these data the receiver (and now verifier) can extract two properties. First, that the data packet is created by a well configured trusted platform. This is done by inspecting the transmitted AIK certificate. Secondly, that the device is in a trustworthy state. This means that the device, hard- and software, has not been altered and is proved by the transmitted trust measurement. The verifier can vet the trust measurement by comparing the measurement with an additionally provided log how the measurement was computed. In this log every component of the device is recorded with its own integrity value, e.g. a hash of a firmware or software. The verifier has to know the reference value to validate the log. A malicious device can modify the log but not tamper the signed measurement. Knowledge of all possible reference values is a serious problem in the PC domain as there are many components with various firmware versions. In the mobile domain the situation is different as there are not so much devices and firmware updates are very rare. Moreover a special working group is standardising a TPM for the mobile domain. This mobile TPM includes a tamper proof built-in verifier. As an alternative to remote attestation addressing privacy concerns, the TCG has defined the Direct Anonymous Attestation (DAA). As DAA requires much higher computing power than remote attestation it has no practical relevance today.

Introducing TC in the context of a VoIP signature device enables to fulfil the desired security requirements by implementing the signature device as a trusted platform. The resulting trusted voice signature terminal would rely on the following base technologies.

- Certificates and an appropriate infrastructure are used to authenticate a communication party in the context of a voice transaction and to link a public key to this party. It also reveals who issued this certificate. In the context of TC, certificates are used to establish trust in the instantiation of a trusted platform. To testify certificates it is necessary to provide an infrastructure providing a root of trust to verify the origin of a certificate.

- To testify the integrity of a single program in the signature device, in this case the signature application, it is necessary to prove the integrity of the runtime environment. Therefore a measurement of the boot process including the involved hardware is required. TC offers the trusted boot process to solve this problem. During boot, every component measures the following before it is started. As the root of this chain, TCG has introduced a special component which measures itself and the BIOS, reports these values to the TPM and loads the BIOS. The BIOS in turn measures, reports, and loads the boot loader, and so on.

- Proving the integrity of the involved devices and storing this information in the voice stream is important for the initialisation of trust between the communication parties. The attestation protocol enables a mutual approval of the integrity of the respective signature devices.

As an extension, a trusted platform can offer a reliable time source which is linked to an external clock provided by a service provider, e.g., a mobile network operator, which enables a trustworthy time stamping by the signature device itself. Thus it is possible to bind a conversation to a certain time without the necessity to request for every conversation an external time stamp.

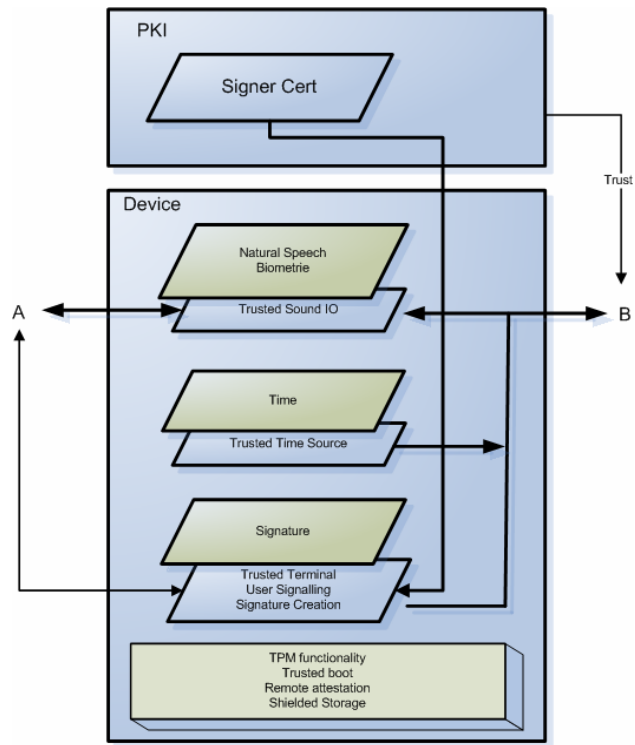

*Figure 4: Schematic presentation of the Trusted Voice Signature Terminal. This device is design as a trusted platform. On this platform trustworthy software can perform several tasks.*

As voice communication is limited to one channel, namely audio, it is not necessary to protect every I/O channel of the signature device. By only protecting the voice channel we assume that the protection level of the envisaged Trusted Voice Signature Terminal is equivalent to the general concept proposed in [22]. This assumption is based on the functionality offered by the signing protocol and the enforcement provided by TC technology. Figure 4 shows a schematic presentation of the envisaged Trusted Voice Signature Terminal. An external PKI infrastructure enables the device to verify third party certificates. The trusted platform offers I/O channels, signature creation, and a reliable time. On top of this basic functions trusted software can perform, e.g., biometry based on the voice of the speaker. The incoming analogue stream is digitised by a verified hardware module and then processed by a trusted software system. This reduces possible attacks on this device to attacks on the analogue audio channel or directly on the TPM

The authentication of a user does not necessarily mean to reveal the identity of the person. Utilising the TC concept of privacy CAs [23] a trusted third party can be added which vouches for the identity of a particular user. This indirection can be used to bind a conversation to an organisation or company which runs the privacy CA.

One should note that using TC in our context differs from the usage in the context of Digital Rights Management (DRM). The economic good, in this case the authentication of the user and attestation of the integrity, is bound to one dedicated TPM. If this particular TPM breaks due to an attack the damage is restricted in time and space. Therefore the possible financial impact is bounded. If a TPM breaks in the DRM scenario a protected good can be changed into an unprotected good and freely distributed.

# 7 CONCLUSIONS

We have shown that it is possible to digitally sign voice transaction in real time and to reach a sufficiently high level of non-repudiation. The inherent security of the concept is high. In particular all integrity checks of Section 5.2 can be applied in real time, thus requiring an attacker to forge natural speech (fitting to context) in real time, which is rather difficult. The ample possibilities to forensically inspect a signed VoIP call enable a dual evaluation of the signed data of the signed content. On the one hand the integrity of the conversation is proved by the electronic signature. On the other hand, natural language contains inherent biometric data, which can be used, e.g., for independent speaker identification, or detection of forgery. Leveraging this synergy raises the evidentiary value of the signed VoIP conversation significantly in comparison to arbitrary, electronically signed, digital documents. As early as 1979, it was foreseen [24 referring to 25] that

> "Digital signatures promise to revolutionize business by phone…"

VoIP as a base technology can finally fulfil that promise through our approach.

The proposed modifications to existing standards are minimal and that this proposal is already interoperable with existing VoIP infrastructures. Using TC as a trust anchor providing a trustworthy infrastructure seems not far fetched as there is much activity in the mobile domain to establish trustworthiness for mobile devices by enhancing the hardware with the mentioned mobile device and special trusted operating systems.

We finally list some research and development goals that remain on the way to a successful, real-world, deployment of signed VoIP. First and foremost, the demonstrated technology must be implemented on mobile phones, preferably equipped with TC technology. Such are scarce as of yet, but the efforts of the mobile industry in the TCG are suggesting that mobile devices with TPMs will be available soon. On this basis, the concept of a trustworthy voice signature device of Section 6 could be realised. Such a device would be able to create signatures over natural language conversations that meet the level of security of qualified digital signatures according to the EU signature directive and, e.g., German law. On the technical side, security policies should be fine-tuned to fit various application domains. In particular, the interplay of QoS threshold, understandability, and interval length deserves further investigation. A legal evaluation should accompany technological development to obtain a sustainable assessment of the probative force and applicability domain of signed digital voice.